\documentclass[a4paper,11pt]{article}
\pdfoutput=1 

\usepackage{amsmath,amssymb}
\usepackage{bm}
\usepackage{graphicx}
\usepackage[section]{placeins}

\usepackage{jheppub} 

\usepackage[T1]{fontenc} 

\title{\boldmath A different kind of continuum limit for the three-dimensional U(1) gauge theory}

\author[a]{Andreas Athenodorou,}
\author[b,c]{Claudio Bonati,}
\author[b,c]{and Ivan Soler Calero}

\affiliation[a]{Computation-based Science and Technology Research Center,\\
The Cyprus Institute, 20 Kavafi Str., Nicosia 2121, Cyprus}
\affiliation[b]{
Dipartimento di Fisica dell'Universit\`a di Pisa, \\
Largo Pontecorvo 3, I-56127 Pisa, Italy}
\affiliation[c]{INFN Sezione di Pisa,\\ Largo Pontecorvo 3, I-56127 Pisa, Italy}

\emailAdd{a.athenodorou@cyi.ac.cy}
\emailAdd{claudio.bonati@unipi.it}
\emailAdd{ivan.soler@df.unipi.it}

\abstract{In three-dimensional compact U(1) lattice gauge theory color
confinement can be understood analytically through the dynamics of magnetic
monopoles. However, its continuum limit is pathological, since the ratio of the
glueball masses to the square root of the string tension vanishes as the
continuum limit is approached.  We investigate a simple extension of the Wilson
action in which the total number of lattice monopoles is coupled to an
additional parameter $\mu$.  By tuning $\beta$ and $\mu$ simultaneously, we
identify a line of constant physics along which the ratio of the lightest
glueball mass to the square root of the string tension remains constant. We
further show that the same scaling is satisfied, within numerical
uncertainties, by the other low-lying glueball masses considered in this work.
Along this trajectory, the string tension in lattice units decreases with
increasing $\beta$, thus suggesting that the modified lattice action may
provide a regularization of three-dimensional compact U(1) gauge theory with a
physically well-behaved continuum limit.}

\begin{document} 
\maketitle
\flushbottom

\section{Introduction}
\label{sec:intro}

Color confinement is a long-distance phenomenon that characterizes the
low-temperature phase of most Yang-Mills theories and requires nonperturbative
methods for its investigation. The lattice formulation~\cite{Wilson:1974sk} has
been extensively used for this purpose, and numerical simulations have provided
overwhelming evidence that the continuum limit of
four-dimensional SU($N$) lattice gauge theories exists and corresponds to a confining quantum field
theory. Confinement can be rigorously established in the strong-coupling regime
of lattice gauge theories~\cite{Wilson:1974sk,Osterwalder:1977pc} (see also,
e.g.,~\cite{Seiler:1982pw, Itzykson:1989sx}), however this proof does not
extend to the weak-coupling regime relevant for the continuum limit. Indeed, a
mathematical proof of confinement in four-dimensional SU($N$) 
Yang-Mills theory remains elusive, and to prove the existence of a
nonvanishing mass gap in this theory constitutes a significant part of the ``Quantum
Yang-Mills Theory'' Millennium Prize Problem~\cite{JaffeWitten2006}.

In this context, a particularly important role is played by the compact
formulation of the three-dimensional U(1) lattice gauge theory. For this theory it
is indeed possible to exploit a dual Coulomb gas representation in terms of
monopole degrees of freedom (which exist only in the compact formulation of the
theory) to prove that confinement persists even in the weak-coupling
regime~\cite{Polyakov:1975rs,Polyakov:1976fu,Gopfert:1981er}, see
also~\cite{Muller:1981ia, Polyakov:1987hqn,Poppitz:2021cxe}. In
the weak-coupling limit the mass gap $m_0$ of the theory, i.e. the mass of the
lightest glueball, is given in units of the lattice spacing $a$ by  
\begin{equation}\label{eq:m0}
am_0=\sqrt{8\pi^2\beta}\, e^{-\pi^2\beta v(0)}\ ,
\end{equation}
where $\beta$ is the dimensionless lattice coupling to be introduced later, 
the weak coupling/con\-tinuum limit corresponds to $\beta\to\infty$, and  
\begin{equation}
v(0)=\int_{-\pi}^{\pi}\frac{\mathrm{d}^3k}{(2\pi)^3} \frac{1}{4\sum_{i=1}^3 \sin^2(k_i/2)}\simeq 0.252731\ldots
\end{equation}
is the lattice Coulomb potential at zero distance. In the same regime the
string tension $\sigma$ can be rigorously shown to satisfy the inequality 
\begin{equation}\label{eq:sigma}
a^2\sigma\ge c_{\sigma}\frac{am_0}{4\pi^2\beta}\ ,
\end{equation}
where $c_{\sigma}$ is positive constant, and the theory is thus confining also for
arbitrarily large $\beta$ values. Moreover, using the semiclassical
approximation, one finds that the actual value of $a^2\sigma$ is given by the
r.h.s. of Eq.~\eqref{eq:sigma} with $c_{\sigma}=8$.

It is thus natural that the three-dimensional compact U(1) lattice gauge theory
has become a paradigmatic model for the investigation of confinement-related
observables, such as the long-distance behavior of the static potential and the
properties of confining flux tubes.  Both these quantities are subject to
strong theoretical constraints coming from Effective String Theory (EST), which
provides an effective description of the low-energy dynamics of the confining
flux tube connecting two static color charges. One of the most remarkable
achievements of EST has been the realization that Lorentz invariance imposes
highly stringent constraints on the form of the effective string action, to the
extent that the string tension is the only free parameter entering the first
few terms of its low-energy expansion~\cite{Luscher:1980fr, Luscher:1980ac,
Luscher:2004ib, Dubovsky:2012sh, Aharony:2013ipa}, a property commonly referred
to as string universality (see, e.g.,~\cite{Brandt:2016xsp,Caselle:2021eir} for
reviews).
The fact that string universality nicely describes
lattice data for $\mathbb{Z}_2$, SU(2) and SU($N$) lattice gauge models
\cite{Caselle:2007yc, Brandt:2010bw, Caselle:2010pf, Athenodorou:2010cs,
Caristo:2021tbk, Caselle:2024zoh} but not those of the compact U(1) gauge
model~\cite{Caselle:2014eka, Caselle:2016mqu} then appears quite surprising.

This puzzling result has been related to a further peculiarity (a pathology, in fact) 
of the compact U(1) model: from Eq.~\eqref{eq:m0} and Eq.~\eqref{eq:sigma} we see that the physical 
ratio $m_0/\sqrt{\sigma}$ is asymptotically equal to (using $c_{\sigma}=8$)
\begin{equation}\label{eq:bad}
\frac{m_0}{\sqrt{\sigma}}=\frac{(2\pi^2\beta)^{3/4}}{\sqrt{2}}e^{-\pi^2\beta v_c(0)/2}\ ,
\end{equation}
and goes to zero in the continuum limit $\beta\to\infty$, see
\cite{Athenodorou:2018sab} for recent numerical results supporting this
behavior. 
This unusual large-$\beta$ scaling behavior was already pointed out
in~\cite{Gopfert:1981er}, and implies that there are two inequivalent ways to take
the continuum limit. One possibility is to keep $m_0$ fixed as the continuum
limit is approached; in this case the string tension $\sigma$
diverges, and the resulting continuum theory is free~\cite{Gopfert:1981er}.
Alternatively, we can keep $\sigma$ fixed while taking the continuum limit, 
and only in this case the application of EST is meaningful. However,
$m_0$ vanishes in the continuum limit, and its contribution to the
low-energy dynamics can no longer be neglected. Indeed, it has recently been
shown that the discrepancies between the predictions of EST and lattice results
are resolved once the effects of $m_0$ are properly incorporated into the
effective theory~\cite{Aharony:2024ctf,Caselle:2025vhx}.

For most lattice models, the continuum limit is associated with the critical
behavior that emerges in the vicinity of a critical point, where all physical
length scales diverge proportionally to the correlation length. This is not the
case for the three-dimensional compact U(1) lattice gauge theory, in which two
independent physical length-scales coexist. In the language of critical
phenomena this case corresponds to a multicritical
point~\cite{LiuFisher73,Fisher:1974zz,Nelson:1974xnq,Kosterlitz:1976zza} (see
also, e.g.,~\cite{Itzykson:1989sx}). To define a continuum limit in which the
ratio of the two diverging length-scales remains finite, one must approach the
multicritical point along a specific trajectory in the space of relevant
couplings. However, in the standard Wilson formulation of the compact three-dimensional U(1)
model we only have a single control parameter, namely $\beta$.

The aim of this work is to investigate whether it is possible to extend the
Wilson action of the compact U(1) lattice gauge theory by introducing an
additional coupling that allows one to approach the continuum limit along a
line of constant physics, namely a trajectory along which the ratio
$m_0/\sqrt{\sigma}$ remains constant. Since the confining properties of the
model are intimately related to the presence of monopoles in its dual
formulation, a natural candidate for the additional interaction is a
monopole-dependent term. To this end, we define lattice monopoles using the
DeGrand-Toussaint construction~\cite{DeGrand:1980eq} and consider an action
supplemented by a term proportional to the total number of lattice monopoles,
with each monopole counted by the absolute value of its magnetic charge. 
To define the line of constant physics, we use the values of the string
tension, extracted from the lightest torelon mass, and of the lightest glueball, namely
the ground state in the $J^{PC}=0^{--}$ channel~\cite{Athenodorou:2018sab}. We
also determine the mass of the first excited $0^{--}$ glueball and that of the
$0^{++}$ glueball ground state, in order to verify that all these masses scale
proportionally to $\sqrt{\sigma}$ as the continuum limit is approached along
the line of constant physics.

The paper is organized as follows. In Sec.~\ref{sec:model}, we introduce the modified
action for the three-dimensional compact U(1) lattice gauge theory and provide
a heuristic motivation for its form. We then describe the simulation algorithm
and the numerical techniques used to determine the string tension and the
glueball spectrum. In Sec.~\ref{sec:results}, we present and discuss our numerical
results. Finally, in Sec.~\ref{sec:concl}, we summarize our findings and outline
possible directions for future work.

\section{The lattice model and numerical methods}
\label{sec:model}

\subsection{The lattice action and the update algorithm}\label{sec:action}

The dynamical degrees of freedom of the compact formulation of the
three-dimensional U(1) lattice gauge theory are complex link variables $U_{{\bm
x},\nu}$ of unit modulus ($|U_{{\bm x},\nu}|=1$), associated with each link of
a three-dimensional cubic lattice, where ${\bm x}$ labels the lattice site and
$\nu=1,2,3$ denotes the link direction. The Euclidean lattice action that will
be adopted in this work is (periodic boundary conditions are always used)
\begin{equation}\label{eq:action}
S=-\beta\sum_{{\bm x},\rho>\nu}\mathrm{Re}P_{{\bm x},\rho\nu} - \mu N_m\ ,
\end{equation}
where $P_{{\bm x},\rho\nu}$ is the plaquette operator,
\begin{equation}\label{eq:plaq}
P_{{\bm x},\rho\nu}=U_{{\bm x},\rho}U_{{\bm x+\hat{\rho}},\nu}
U^{\dag}_{{\bm x+\hat{\nu}},\rho}U^{\dag}_{{\bm x},\nu},
\end{equation}
and $N_m$ denotes the total monopole number, defined according to the
DeGrand-Toussaint prescription~\cite{DeGrand:1980eq} (summarized in
App.~\ref{sec:mono}) as the sum of the absolute values of the monopole charges
$q_{\bm x}$,
\begin{equation}\label{eq:Nm}
N_m=\sum_{\bm x} |q_{\bm x}|\ .
\end{equation}
The first term in Eq.~\eqref{eq:action} is the standard Wilson
action~\cite{Wilson:1974sk}. In the \emph{naive} continuum limit, the
dimensionless lattice coupling $\beta$ is related to the dimensionful gauge
coupling $g$ by $\beta=1/(g^2a)$. The additional term is introduced to provide
a second tunable coupling, making it possible to approach the continuum limit
along a line of constant physics.

To gain an intuitive understanding of why the second term in
Eq.~\eqref{eq:action} can be used to keep the ratio $m_0/\sqrt{\sigma}$ constant,
we can examine how it modifies Polyakov's classical analysis of
the model, see, e.g.,~\cite{Polyakov:1987hqn} \S 4-5.
By decomposing the gauge field into a topologically trivial component and a
monopole contribution, the lattice partition function can be factorized into a Gaussian
(spin-wave) part and a monopole part. The latter is governed by the effective
action
\begin{equation}
S_{\mathrm{mono}}=\frac{2\pi}{4g^2}\sum_{\bm{x}\neq {\bm y}} \frac{q_{\bm x}q_{\bm y}}{|{\bm x}-{\bm y}|}
+\frac{\mathrm{const}}{g^2}\sum_{\bm x}q_{\bm x}^2-\mu \sum_{\bm x}|q_{\bm x}|\ ,
\end{equation}
where the last term is precisely the contribution introduced in Eq.~\eqref{eq:action}.
An approximate solution is then obtained by restricting the monopole charges to
$q_{\bm x}=\pm 1$ and assuming a dilute monopole gas. Under these assumptions,
the monopole partition function can be rewritten in terms of an auxiliary field
$\chi({\bm x})$, whose dynamics is described by the sine-Gordon action
\begin{equation}
S_{\mathrm{sG}}=\left(\frac{g}{2\pi}\right)^2\int \Big[\big(\nabla \chi({\bm x})\big)^2-
M^2\cos\big(\chi({\bm x})\big)\Big]\mathrm{d}{\bm x}\ ,
\end{equation}
where the squared mass $M^2$ is proportional (up to subleading corrections in $\beta$) to the monopole fugacity
\begin{equation}
\zeta=\exp(-\mathrm{const}/g^2+\mu)\ .
\end{equation}
The parameter $M$ is, in turn, proportional both to the mass gap $m_0$ and to
the string tension $\sigma$, see Eqs.~\eqref{eq:m0} and~\eqref{eq:sigma} (for
$\mu=0$). It follows that keeping the ratio $m_0/\sqrt{\sigma}$ fixed requires
the monopole fugacity $\zeta$ to remain approximately constant, implying the scaling
$\mu\propto 1/g^2\propto \beta$.  It should be noted, however, that in the
standard case $\mu=0$ one has $\zeta\ll 1$ as $\beta\to\infty$, so that the
dilute-gas approximation is self-consistent. In contrast, if $\zeta$ is kept
fixed while taking the continuum limit, there is no reason to expect the
monopole gas to remain dilute. In fact, a priori, we cannot even be sure that
a continuum limit exists at all.

Gauge configurations are sampled according to the probability distribution
$e^{-S}/Z$, with the action defined in Eq.~\eqref{eq:action}, using an
overrelaxed algorithm~\cite{Adler:1981sn} that combines Metropolis and
pseudo-microcanonical updates. In the Metropolis step~\cite{Metropolis:1953am},
we propose the change $U_{{\bm x},\nu}\to e^{i\theta}U_{{\bm x},\nu}$, where
$\theta$ is drawn uniformly from the interval $[-\epsilon,\epsilon]$. The
proposed move is then accepted with probability $\min(1,e^{-\Delta S})$. The
parameter $\epsilon$ is tuned at the beginning of each simulation to yield an
acceptance rate of approximately $50\%$.  Let us denote by $\Sigma_{{\bm x},\nu}$ the
sum of the staples attached to the link $U_{{\bm x},\nu}$. For $\mu=0$, the
transformation~\cite{Creutz:1987xi}
\begin{equation}\label{eq:micro}
U_{{\bm x},\nu}\to U^*_{{\bm x},\nu}
\left(\frac{\Sigma^{*}_{{\bm x},\nu}}{|\Sigma_{{\bm x},\nu}|}\right)^2\,,
\end{equation}
is an exact microcanonical update: it is involutive and leaves the Wilson
action invariant. For $\mu\neq 0$, we use the same transformation as a proposal,
and accept or reject it with a Metropolis test. The resulting pseudo-microcanonical
update has a typical acceptance rate of approximately $90\%$.
Each update cycle consists of one Metropolis sweep over the whole lattice, followed
by ten pseudo-microcanonical sweeps. 
Measurements are performed every 50 update cycles and for each pair of $\beta$
and $\mu$ values considered we collect approximately $5\times10^4$
measurements.

\subsection{The computation of the U(1) spectrum}\label{sec:spectrum}

In this section, we briefly summarize the methodology employed to determine the
U(1) spectrum. In particular, we describe the calculation of the ground- and
first-excited-state glueball masses in the $0^{--}$ channel, together with the
ground-state mass in the $0^{++}$ channel. We also present the extraction of
the torelon (flux-tube) masses, from which the string tension is subsequently
determined.

\subsubsection{Calculation of glueball masses}\label{sec:spectrum_glueballs}

We begin our discussion of the U(1) spectrum by focusing on the determination
of the glueball masses, noting that the same methodology is employed for the
extraction of the torelon masses. For a more detailed account of the
corresponding calculations in lattice U(1) gauge theory, we refer the reader to
Ref.~\cite{Athenodorou:2018sab}.

The extraction of the spectrum in lattice gauge theories relies on the
computation of Euclidean two-point correlation functions,
\begin{equation}
  C(t)
  =
  \langle \Phi^\dagger(t)\Phi(0)\rangle
  =
  \sum_n |\langle n|\Phi|vac \rangle|^2 \exp\{-E_nt\}
  \stackrel{t\to\infty}{=}
  |\langle 0|\Phi|vac \rangle|^2 \exp\{-aE_0n_t\},
\label{eq:eqn_Mcor}
\end{equation}
where the interpolating operators $\Phi(t)$ are gauge singlets constructed to
transform according to the symmetries of the theory and to carry the quantum
numbers of the states whose energies are to be determined; for the case of
glueballs $J^{PC}$. In the limit of large Euclidean time separations, the
contributions from excited states are exponentially suppressed, and the
correlator becomes dominated by the ground state $E_{0}$ with the appropriate
quantum numbers. The interpolating operators $\Phi(t)$ are projected onto zero
spatial momentum by summing over all spatial positions, thereby enforcing
translational invariance. Consequently, the energies extracted from the
correlation functions correspond directly to the masses of the physical states
i.e. $E_{J^{PC}} = M_{J^{PC}}$.

In lattice simulations, the correlation function in Eq.~\eqref{eq:eqn_Mcor} is
estimated by averaging over an ensemble of gauge-field configurations generated
using Monte Carlo methods. While the statistical uncertainty of the correlator
remains approximately independent of the Euclidean time separation $t$, its
expectation value decreases exponentially with increasing $t$. As a result, the
signal-to-noise ratio deteriorates rapidly at large temporal separations,
making it challenging to isolate the asymptotic regime from which the physical
energies are extracted. It is therefore crucial to construct interpolating
operators with a large overlap onto the desired ground state, thereby
minimizing excited-state contamination and allowing reliable mass
determinations already at relatively small values of $t$.

A standard strategy to improve the overlap with low-lying states is to employ
smeared or blocked operators, which are obtained by iteratively averaging the
gauge links in a gauge-covariant manner~\cite{Athenodorou:2018sab}. These
operator-improvement techniques significantly enhance the projection onto the
low-energy spectrum and lead to a substantial improvement in the quality of the
extracted masses.

In practice, to extract the ground state from Eq.~\eqref{eq:eqn_Mcor} we study
the effective energy $E_{\rm eff}(n_t)$:
\begin{equation}
     aE_{\rm eff}(n_t) = \ln\left(
\frac{C(n_t)}{C(n_t+1)}
\right)\,,
\label{eq:effective_energy}
\end{equation}
as a function of the Euclidean time separation $t=a n_t$. The appearance of a
plateau indicates that the correlation function is dominated by a single
exponential, implying that contributions from excited states have become
negligible. The ground-state energy is then extracted by performing a fit over
the plateau region. To properly account for the statistical correlations
between different time slices, the jackknife resampling method is employed,
following the approach used in previous studies~\cite{Athenodorou:2018sab}. For
lattices with periodic boundary conditions in the temporal direction, the
exponential behaviour is replaced by the corresponding hyperbolic cosine form
to account for backward-propagating states.

Since the statistical uncertainty of $E_{\rm eff}(n_t)$ increases exponentially
with $n_t$, precise determinations of masses and energies require both a
sufficiently light state and an interpolating operator with a large overlap
onto the corresponding physical state. This motivates the construction of
optimized operator bases, which maximize the overlap with the low-lying
spectrum while suppressing excited-state contamination, and therefore
constitute an essential ingredient of high-precision lattice spectroscopy.

From the discussion above, it is evident that the effective energy extracted
from a single correlation function provides direct access only to the ground
state. To determine the excited-state spectrum, one must instead employ a basis
of interpolating operators and solve the \emph{Generalized Eigenvalue Problem}
(GEVP)~\cite{Luscher:1990ck,Blossier:2009kd,Morningstar:1999rf}.

Rather than constructing the correlation function of Eq.~\eqref{eq:eqn_Mcor}
from a single interpolating operator, one considers a basis of operators
carrying the same quantum numbers and forms the corresponding correlation
matrix,
\begin{equation}
    C_{ij}(t)=\langle \Phi_i^\dagger(t)\Phi_j(0)\rangle.
\end{equation}
The interpolating operators have different overlaps with the physical states,
providing a basis capable of resolving the low-lying spectrum. In the case of
glueballs, the operator basis consists of linear combinations of closed Wilson
loops of various shapes and orientations, combined such that the resulting
operators transform irreducibly under the lattice symmetry group with the
desired quantum numbers $J^{PC}$. In this work, we employ the operator basis
introduced in Ref.~\cite{Athenodorou:2016ebg}, constructed at several blocking
levels to improve the overlap with the low-lying states. The energies are
subsequently extracted by solving the generalized eigenvalue problem.

The eigenvectors resulting from GEVP determine the optimal linear combinations
of interpolating operators that maximize the overlap with the individual
physical states. In particular, the eigenvector associated with the largest
eigenvalue has the largest overlap with the ground state, while the remaining
eigenvectors project onto the excited states in order of increasing energy. The
energies $E_n$ are subsequently obtained by constructing effective energies
from the generalized eigenvalues, in complete analogy with
Eq.~\eqref{eq:effective_energy}. In this way, both the ground state and its
excitations can be determined in a systematic manner.

\subsubsection{Calculation of torelon masses and string tension} \label{sec:spectrum_torelons_string}

The confining string tension is determined from the energy of a closed flux
tube, or \emph{torelon}, winding around one of the periodic spatial directions
of the lattice. Without loss of generality, we consider the $x$-direction and
denote its spatial extent by $l=aL$. The corresponding ground-state energy of a
flux tube winding once around this compact direction is denoted by $E_f(l)$.

The numerical procedure closely parallels that employed for the glueball
spectrum. The principal difference lies in the construction of the
interpolating operators. Instead of contractible closed Wilson loops, we
consider gauge-invariant operators built from closed but topologically
non-trivial paths that wind around the spatial torus. These winding operators
are simply spatial Polyakov loops. As in the glueball analysis, we employ
blocking and smearing techniques to improve the overlap with the low-lying
physical states and solve the GEVP to optimize the operator basis. The
resulting principal correlators yield the spectrum of closed flux tubes, from
which the ground-state flux-tube energy $E_f(l)$ is extracted. Although the
GEVP provides access to both the ground state and the excited-state spectrum,
in this work we restrict our analysis to the ground state, since it is
sufficient for the determination of the string tension. The investigation of
the excited flux-tube spectrum is deferred to future work. The operator basis
used to construct the correlation matrix is the same as that employed in
Ref.~\cite{Athenodorou:2011rx}.

For sufficiently long flux tubes, the energy is expected to grow linearly with its length,
\begin{equation}
    E_f(l)\simeq \sigma l,
\end{equation}
where $\sigma$ is the confining string tension. In principle, $\sigma$ could
therefore be obtained directly from the slope of the linear dependence. In
practice, however, simulations at very large values of $l$ become increasingly
demanding, since the corresponding correlation functions decay more rapidly and
the statistical precision required to determine $E_f(l)$ increases
substantially. As a result, the string tension is extracted from flux tubes of
moderate length, for which corrections to the leading linear behaviour must be
taken into account.

To describe these corrections, we fit the measured flux-tube energies using the
Nambu-Goto (GGRT~\cite{Goddard:1973qh}) prediction,
\begin{equation}
   E_f(l)=\sigma l
   \left(
   1-\frac{\pi}{3\sigma l^2}
   \right)^{1/2},
   \label{eq:NG}
\end{equation}
which resums the universal finite-length corrections to the classical linear
behaviour~\cite{Aharony:2013ipa}. This expression has been extensively tested
and found to accurately describe flux-tube energies in a wide range of
confining gauge theories in $D=2+1$, as well as in $D=3+1$, including
$SU(N)$~\cite{Athenodorou:2011rx}, compact $U(1)$~\cite{Athenodorou:2018sab},
and $Z_N$~\cite{Luo:2023cjv} gauge theories, even for relatively short flux
tubes characterized by small values of $l\sqrt{\sigma}\simeq 1-2$.

The applicability of Eq.~\eqref{eq:NG} further requires that finite-volume
effects associated with the transverse directions remain negligible. In
particular, the transverse lattice extent, $l_\perp=a  L_{\perp}$, must be
sufficiently larger than the flux-tube length to suppress interactions with its
periodic contributions in the spectrum. To verify that this condition is
satisfied for our simulations, we performed a dedicated finite-volume study.
The details and results of this analysis are presented in
Section~\ref{sec:results}.

\section{Numerical results}
\label{sec:results}

\begin{table}[t]
  \centering
  \begin{tabular}{l|l|l|l}
  $\beta$ & $\mu$ & $L_{\perp}$ & torelon mass \\ \hline
  2.0  & 0.3 & 18 & 0.6897(30)  \\ \hline
  2.0  & 0.3 & 24 & 0.6869(17)  \\ \hline\hline
  2.0  & 0.5 & 18 & 0.8118(42) \\ \hline
  2.0  & 0.5 & 24 & 0.8093(22) \\ \hline\hline
  2.1  & 0.5 & 22 & 0.5760(27)  \\ \hline
  2.1  & 0.5 & 30 & 0.5794(38) \\ \hline\hline
  2.1  & 0.7 & 22 & 0.6775(38)  \\ \hline
  2.1  & 0.7 & 30 & 0.6768(48)  
  \end{tabular}
  \caption{Finite-volume study of the torelon mass used to extract the string
  tension. The string tension is measured on lattices of size $12\times L_{\perp} \times
  40$ at $\beta=2.0$ and $12\times L_{\perp} \times 48$ at $\beta=2.1$.}
  \label{tab:finite_volume}
\end{table}

To determine a line of constant physics for the three-dimensional compact U(1)
lattice gauge theory with action~\eqref{eq:action}, we use the string tension
and the mass of the lightest glueball, corresponding to the ground state of the
$0^{--}$ channel and denoted by $M_{0^{--}}^{\mathrm{gs}}$. As a consistency
check, we also determine the mass of the first excited state in the same
channel, $M_{0^{--}}^{\mathrm{ex1}}$, and that of the ground state in the
$0^{++}$ channel, $M_{0^{++}}^{\mathrm{gs}}$.

The spectrum of the compact U(1) lattice gauge theory with Wilson action
(corresponding to the case $\mu=0$) has been investigated in \cite{Athenodorou:2018sab},
and to fix the starting point of the line of constant physics we use the values
reported in Tab.~5 of that paper for $\beta=1.8$:\footnote{The choice of $\beta=1.8$ at $\mu=0$ 
is obviously quite arbitrary, with different choices for the starting point 
corresponding to different lines of constant physics.}
\begin{equation}
aM_{0^{--}}^{\mathrm{gs}}=0.6996(21)\ ;\quad aM_{0^{++}}^{\mathrm{gs}}=1.194(11)\ ;\quad a\sqrt{\sigma}=0.3019(18)   
\end{equation}
from which we get 
\begin{equation}\label{eq:ratios_start}
M_{0^{--}}^{\mathrm{gs}}/\sqrt{\sigma}=2.317(15) \ ;\quad M_{0^{++}}^{\mathrm{gs}}/\sqrt{\sigma}=3.955(43)\ .
\end{equation}
To determine the line of constant physics, for each value of $\beta$ larger
than 1.8 considered we perform simulations at several values of $\mu$, and
interpolate the results to obtain the value $\mu(\beta)$ satisfying
\begin{equation}\label{eq:cpc}
M_{0^{--}}^{\mathrm{gs}}\big(\beta,\mu(\beta)\big)/\sqrt{\sigma\big(\beta,\mu(\beta)\big)}=2.317 \ . 
\end{equation}

\begin{figure}[b]
  \centering
  \includegraphics*[width=0.45\columnwidth]{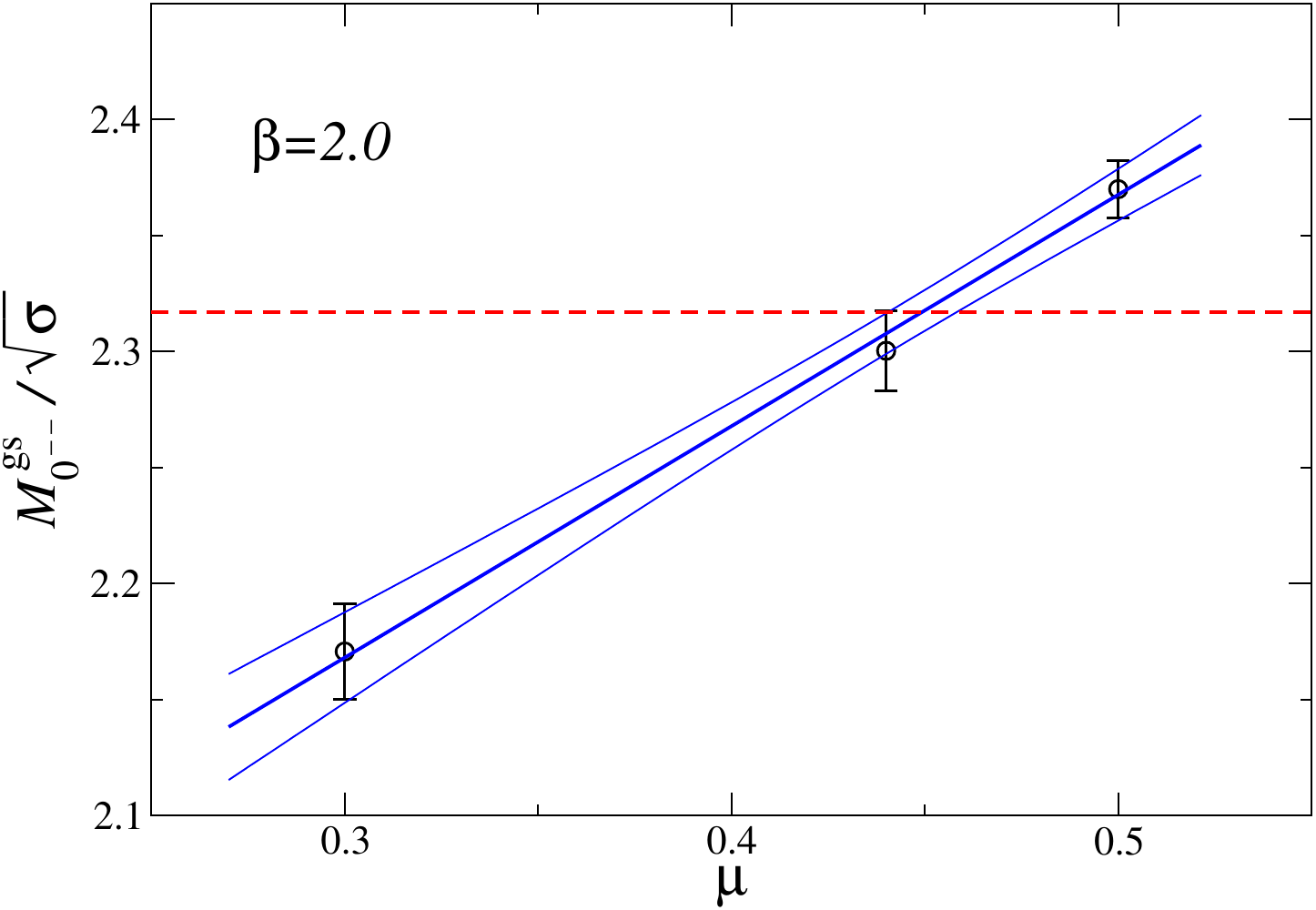}
  \qquad
  \includegraphics*[width=0.45\columnwidth]{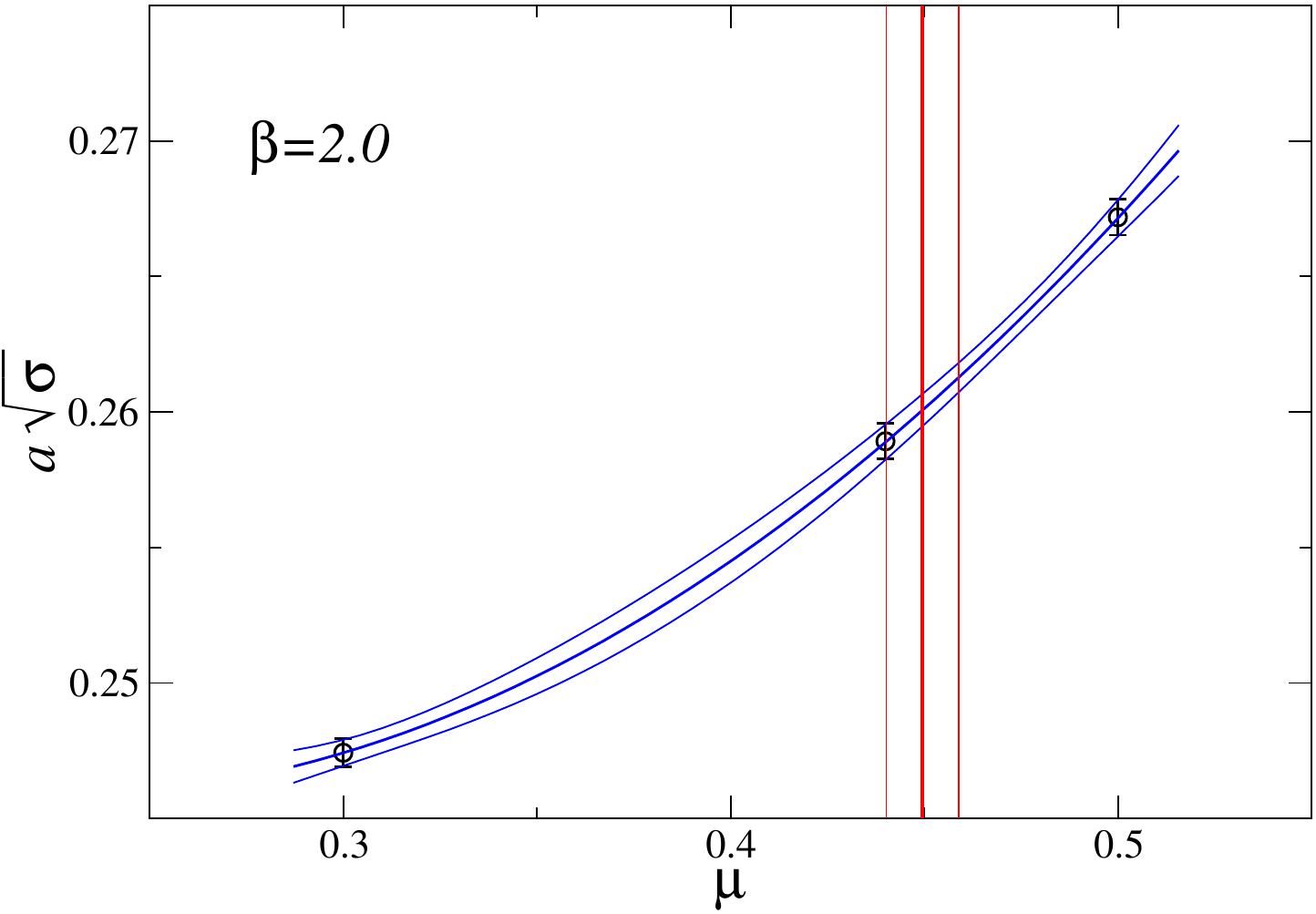}
  \caption{(left) Behavior of $M_{0^{--}}^{\mathrm{gs}}$/$\sqrt{\sigma}$ as a
  function of $\mu$ for $\beta=2$ (see Tab.~\ref{tab:res2.0}). The blue band
  represents the one-standard-deviation uncertainty of the linear fit, while the
  horizontal dashed line indicates the value given in Eq.~\eqref{eq:cpc}.
  (right) Behavior of $a \sqrt{\sigma}$ as a
  function of $\mu$ for $\beta=2$ (see Tab.~\ref{tab:rawres2.0}). The blue band
  represents the one-standard-deviation uncertainty of the quadratic fit, while the
  vertical red band indicates the value $\mu(\beta)$ obtained by imposing Eq.~\eqref{eq:cpc}.
  }
  \label{fig:m0sigma_beta2}
\end{figure}

Glueball masses have been determined using the procedure described in
Sec.~\ref{sec:spectrum}, on symmetric lattices of the same size as those
employed in~\cite{Athenodorou:2018sab} at the corresponding values of $\beta$.
Since all glueball masses are observed to increase in lattice units as $\mu$ is
increased (see Tabs.~\ref{tab:rawres1.9}-\ref{tab:rawres2.2} in
App.~\ref{sec:rawdata}), finite-volume effects are expected to be well under
control for these observables.
To determine the string tension, we used asymmetric lattices for all values of
$\beta$ except $\beta=1.9$. Dedicated finite-volume tests, see Table~\ref{tab:finite_volume},
confirm that finite-volume effects are well under control also in this case.

\begin{table}[h!]
  \centering
  \begin{tabular}{l|l|l|l}
  $\mu$ & $M_{0^{--}}^{\mathrm{gs}}$/$\sqrt{\sigma}$  &  
          $M_{0^{--}}^{\mathrm{ex1}}$/$\sqrt{\sigma}$ &  
          $M_{0^{++}}^{\mathrm{gs}}$/$\sqrt{\sigma}$ \\ \hline
  0$^*$    &   2.131(10) & 5.197(31)  &  3.779(43)  \\  \hline  
  0.1  &   2.222(27) & 5.30(12)   &  3.963(79)  \\ \hline
  0.2  &   2.294(33) & 5.56(23)   &  3.944(92)   
  \end{tabular}
  \caption{Mass ratios in units of the square root of the string tension 
  for $\beta=1.9$ and several values of $\mu$
  (for the corresponding raw data see App.~\ref{sec:rawdata}).
  For $\mu=0$, we quote the results of~\cite{Athenodorou:2018sab}.}
  \label{tab:res1.9}
\end{table}

\begin{table}[h!]
  \centering
  \begin{tabular}{l|l|l|l}
  $\mu$ & $M_{0^{--}}^{\mathrm{gs}}$/$\sqrt{\sigma}$  &  
          $M_{0^{--}}^{\mathrm{ex1}}$/$\sqrt{\sigma}$ &  
          $M_{0^{++}}^{\mathrm{gs}}$/$\sqrt{\sigma}$ \\ \hline
  0.3  &  2.171(21) &  5.274(41) & 3.830(70) \\  \hline  
  0.44 &  2.300(17) &  5.400(75) & 3.986(29) \\ \hline
  0.5  &  2.370(12) &  5.449(42) & 3.948(49)  
  \end{tabular}
  \caption{Mass ratios in units of the square root of the string tension 
  for $\beta=2.0$ and several values of $\mu$
  (for the corresponding raw data see App.~\ref{sec:rawdata}).}
  \label{tab:res2.0}
\end{table}

\begin{table}[h!]
  \centering
  \begin{tabular}{l|l|l|l}
  $\mu$ & $M_{0^{--}}^{\mathrm{gs}}$/$\sqrt{\sigma}$  &  
          $M_{0^{--}}^{\mathrm{ex1}}$/$\sqrt{\sigma}$ &  
          $M_{0^{++}}^{\mathrm{gs}}$/$\sqrt{\sigma}$ \\ \hline
  0.5  &  2.114(20) &  5.06(11)  & 3.766(56) \\  \hline  
  0.7  &  2.295(22) &  5.447(44) & 3.992(49) \\ \hline
  0.9  &  2.475(12) &  5.61(13)  & 4.053(36)  
  \end{tabular}
  \caption{Mass ratios in units of the square root of the string tension 
  for $\beta=2.1$ and several values of $\mu$
  (for the corresponding raw data see App.~\ref{sec:rawdata}).}
  \label{tab:res2.1}
\end{table}

\begin{table}[h!]
  \centering
  \begin{tabular}{l|l|l|l}
  $\mu$ & $M_{0^{--}}^{\mathrm{gs}}$/$\sqrt{\sigma}$  &  
          $M_{0^{--}}^{\mathrm{ex1}}$/$\sqrt{\sigma}$ &  
          $M_{0^{++}}^{\mathrm{gs}}$/$\sqrt{\sigma}$ \\ \hline
  1     & 2.319(11) &  5.434(59)    &  4.010(60) \\ \hline
  1.1   & 2.427(11) &  5.458(65)    & 3.992(64) 
  \end{tabular}
  \caption{Mass ratios in units of the square root of the string tension 
  for $\beta=2.2$ and two values of $\mu$
  (for the corresponding raw data see App.~\ref{sec:rawdata}).}
  \label{tab:res2.2}
\end{table}

\begin{table}[h!]
  \centering
  \begin{tabular}{l|l|l|l|l}
   $\beta$ & 
   $\mu(\beta)$ & 
   $M_{0^{--}}^{\mathrm{ex1}}/\sqrt{\sigma}$ & 
   $M_{0^{++}}^{\mathrm{gs}}/\sqrt{\sigma}$ & 
   $a\sqrt{\sigma}$ \\ \hline
   1.9   &   0.230(45)  & 5.40(20)  & 4.02(10)   & 0.2802(80) \\ \hline
   2.0   &   0.4495(93) & 5.405(31) & 3.9674(25) & 0.2602(14) \\ \hline
   2.1   &   0.725(11)  & 5.461(43) & 3.952(27)  & 0.2479(13) \\ \hline
   2.2   &   0.997(10)  & 5.433(62) & 4.014(62)  & 0.2355(12) 
   \end{tabular}
   \caption{Values of $\mu(\beta)$, $M_{0^{--}}^{\mathrm{ex1}}/\sqrt{\sigma}$,
   $M_{0^{++}}^{\mathrm{gs}}/\sqrt{\sigma}$  and $a\sqrt{\sigma}$ along the line
   of constant physics. These values have been obtained by interpolation as
   discussed in the text.}
   \label{tab:final}
\end{table}

Results obtained for the glueball masses normalized with the square root of the
string tension are reported in Tabs.~\ref{tab:res1.9}-\ref{tab:res2.2}, while
raw lattice data are reported in App.~\ref{sec:rawdata}. We verified that in
all cases a linear fit of $M_{0^{--}}^{\mathrm{gs}}/\sqrt{\sigma}$ data is
sufficient to extract the value $\mu(\beta)$ corresponding to the condition in
Eq.~\eqref{eq:cpc} and the error on $\mu(\beta)$ has been evaluated by
bootstrap, see Fig.~\ref{fig:m0sigma_beta2} for an example of this procedure.
Once the value of $\mu(\beta)$ has been determined for a given $\beta$, the
same fitting and bootstrap procedure is used to interpolate the ratios
$M_{0^{--}}^{\mathrm{ex1}}/\sqrt{\sigma}$ and
$M_{0^{++}}^{\mathrm{gs}}/\sqrt{\sigma}$, as well as the string tension
$a\sqrt{\sigma}$ in lattice units. A linear fit is found to be sufficient for
both glueball-mass ratios, whereas a quadratic interpolation is used for
$a\sqrt{\sigma}$ (see Fig.~\ref{fig:m0sigma_beta2}) at all values of $\beta$
except $\beta=2.2$, where only two values of $\mu$ are available. The results
of this analysis are reported in Tab.~\ref{tab:final}.

\begin{figure}[t]
  \centering
  \includegraphics*[width=0.45\columnwidth]{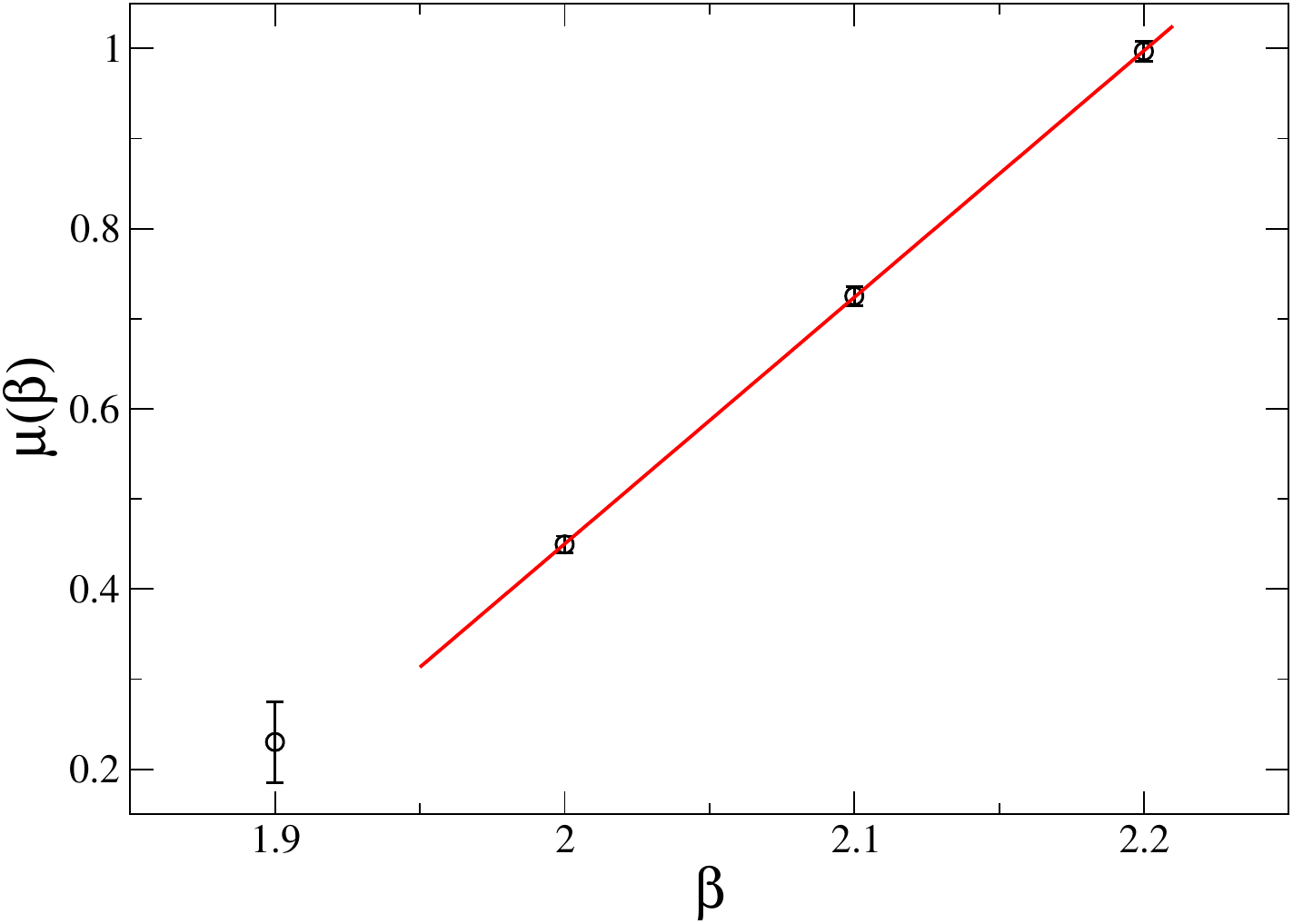}
  \caption{Behavior of $\mu(\beta)$ as a function of $\beta$ along the line of
  constant physics. The solid line shows a linear fit to the three
  largest-$\beta$ data points.}
  \label{fig:mu_beta}
\end{figure}

Fig.~\ref{fig:mu_beta} shows the dependence of $\mu(\beta)$, which defines
the line of constant physics, on the lattice coupling $\beta$. The observed
dependence is approximately linear, in agreement with the \emph{naive}
analytical expectations discussed in Sec.~\ref{sec:action}. A linear fit to the
three largest-$\beta$ data points yields
\begin{equation}
\mu(\beta)=-5.02(15)+2.737(71)\beta,
\end{equation}
whose slope is remarkably close to the value $\pi^2v(0) \approx 2.49$ predicted
under the assumption of a dilute monopole gas (see Eq.~\eqref{eq:m0}).
Note however that, on the contrary of what happens in the standard
compact U(1) theory for $\mu=0$, monopoles are not generically expected to be
diluted. We will came back to this point at the end of the paper.

\begin{figure}[b]
  \centering
  \includegraphics*[width=0.45\columnwidth]{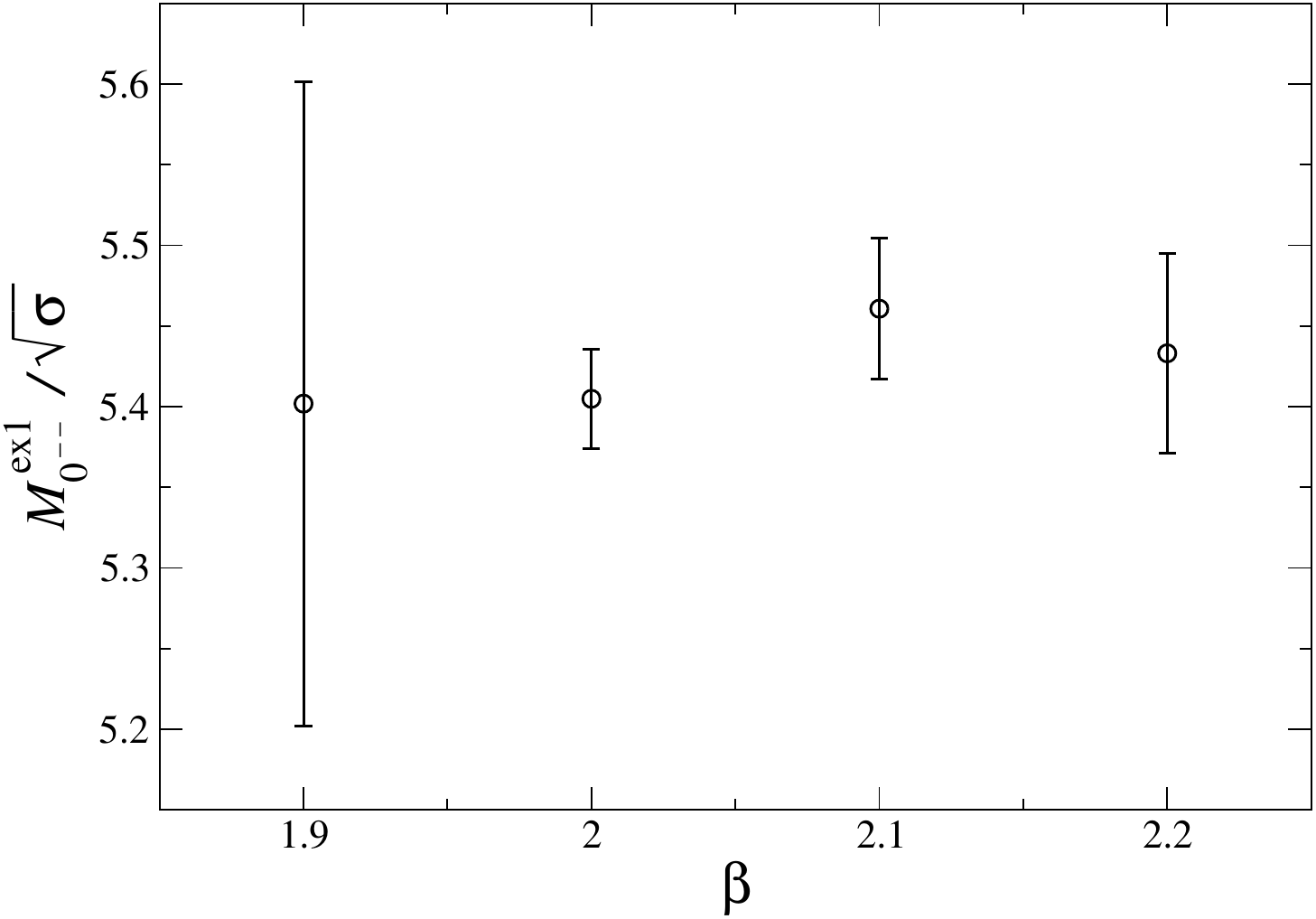}
  \qquad
  \includegraphics*[width=0.45\columnwidth]{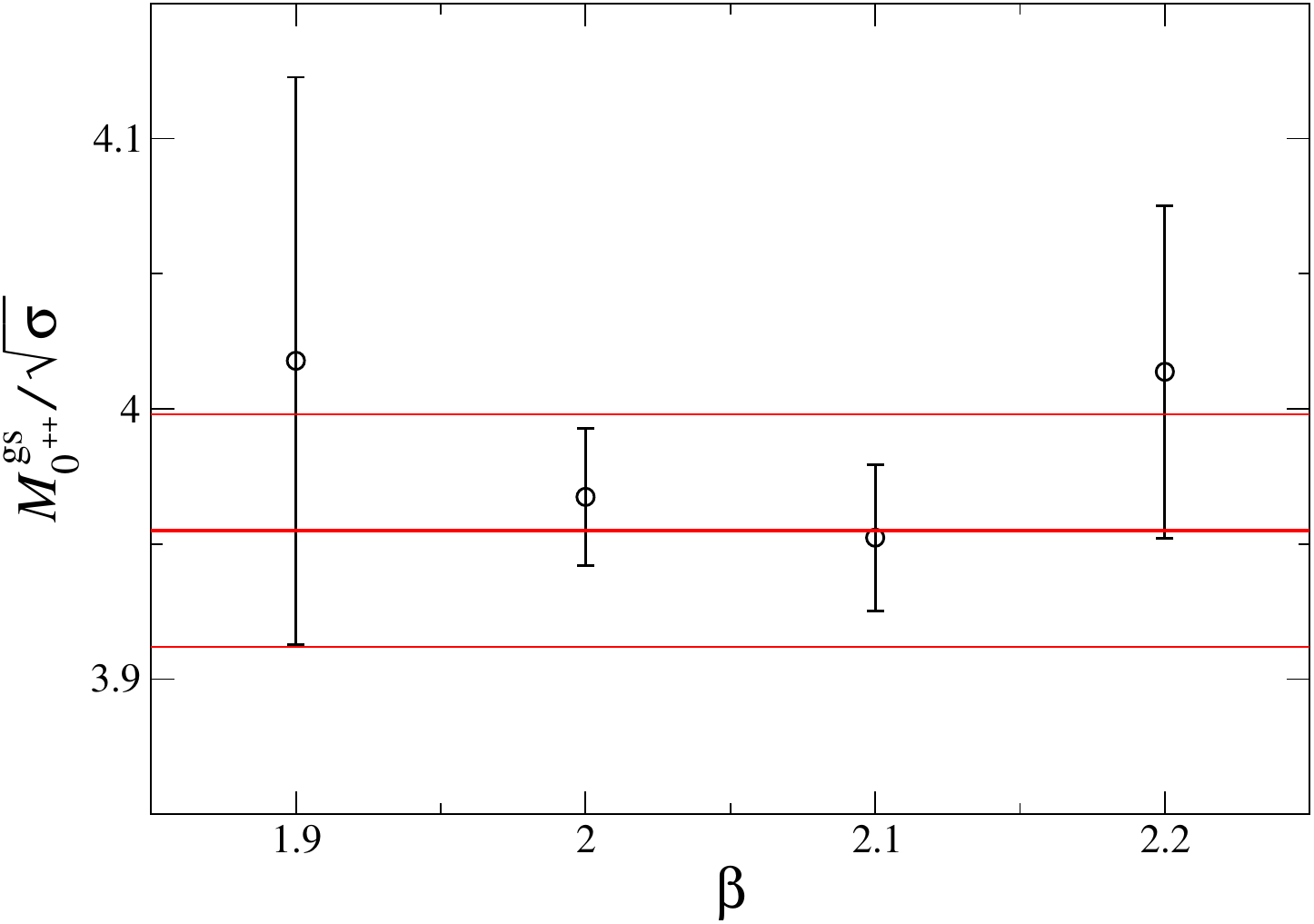}
  \caption{Behavior of the ratios $M_{0^{--}}^{\mathrm{ex1}}/\sqrt{\sigma}$
  (left) and $M_{0^{++}}^{\mathrm{gs}}/\sqrt{\sigma}$ (right) as a function of
  $\beta$ along the line of constant physics. In the right panel the horizontal band 
  denotes the value of $M_{0^{++}}^{\mathrm{gs}}/\sqrt{\sigma}$ estimated
  in~\cite{Athenodorou:2018sab} for $\mu=0$ and $\beta=1.8$, see Eq.~\eqref{eq:ratios_start}.}
  \label{fig:otherM}
\end{figure}

The values of the function $\mu(\beta)$ have so far been determined by requiring
$M_{0^{--}}^{\mathrm{gs}}/\sqrt{\sigma}=\mathrm{const}$, as in
Eq.~\eqref{eq:cpc}. It remains, however, to verify that this condition indeed
defines a line of constant physics, i.e. that all mass ratios are independent
of $\beta$ along this trajectory, at least up to discretization effects.
Numerical results for the ratios $M_{0^{--}}^{\mathrm{ex1}}/\sqrt{\sigma}$ and
$M_{0^{++}}^{\mathrm{gs}}/\sqrt{\sigma}$ are shown in Fig.~\ref{fig:otherM}. 
Although the uncertainties are relatively large, both ratios are consistent
with being independent of $\beta$ as expected along a line of constant
physics. This shows that tuning the parameter $\mu$ does not simply remove the
pathological scaling of the ground-state mass by transferring it to
excited states. Rather, it leads to a consistent line of constant physics for
the low-lying spectrum.

One fundamental issue remains to be addressed, namely whether increasing
$\beta$ along the line of constant physics indeed leads to a continuum limit.
\emph{A priori}, three scenarios are possible: the lattice model has a critical
point at $\beta=\infty$; it has a critical point at some finite value
$\beta=\beta_c$; or it has no critical point at all. In the first two cases, a
nonperturbative continuum limit can be defined by approaching the critical
point while fixing the lattice spacing through a physical observable, for
example by requiring the string tension to remain constant in physical units.
In the third case, instead, the correlation length remains finite throughout the
whole phase diagram (with the possible presence of discontinuous transitions),
preventing the definition of a continuum quantum field theory based on this
lattice model.

\begin{figure}[t]
  \centering
  \includegraphics*[width=0.45\columnwidth]{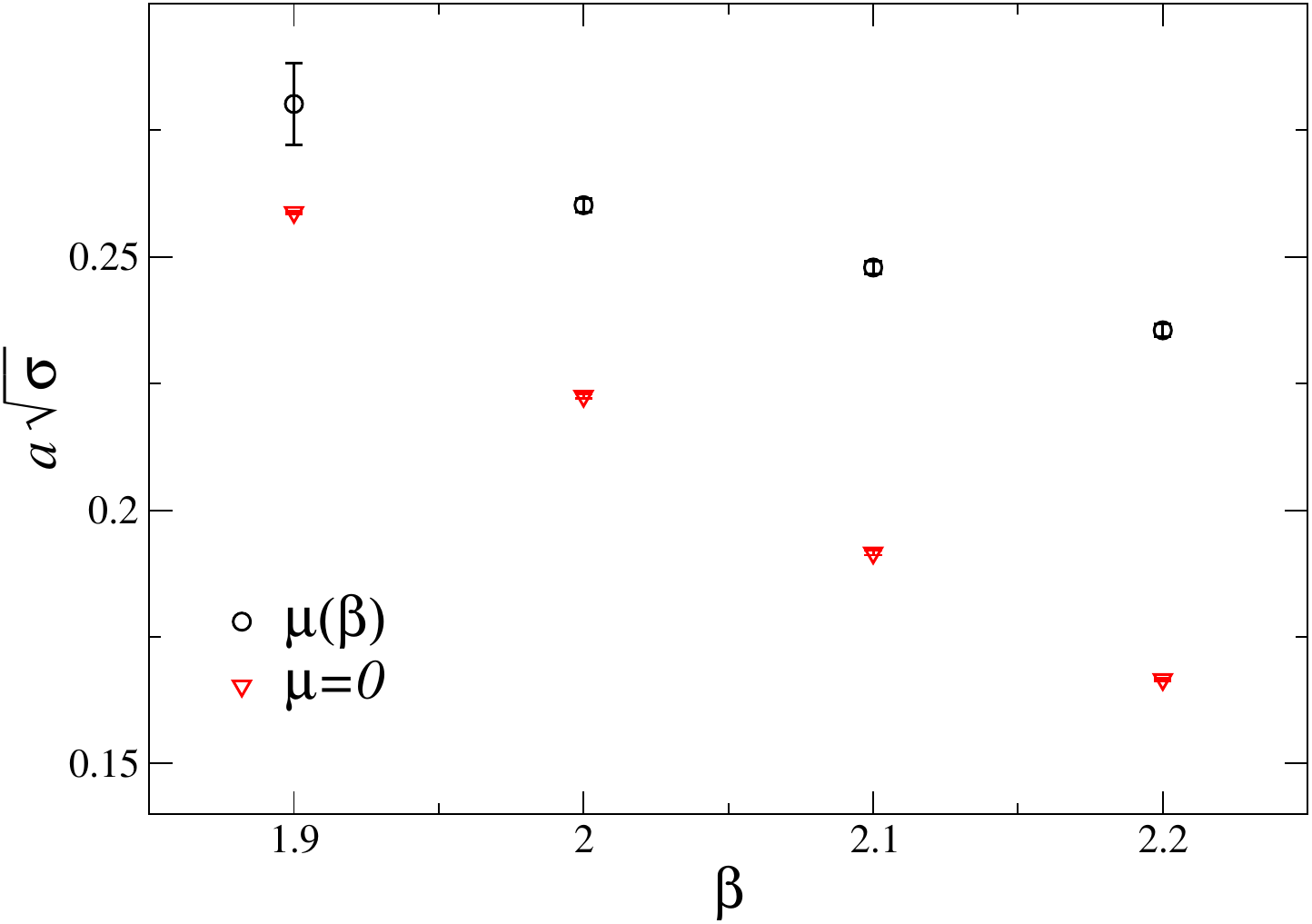}
  \caption{Behavior of the string tension in lattice units as a
  function of $\beta$ along the line of constant physics. For comparison we also report
  the results for $\mu=0$ obtained in~\cite{Athenodorou:2018sab}.
  }
  \label{fig:string}
\end{figure}

In Fig.~\ref{fig:string} we show the string tension in lattice units as a function
of $\beta$ along the line of constant physics. The values of $a\sqrt{\sigma}$
decrease as $\beta$ increases, although the decrease is significantly slower
than that observed for the Wilson action ($\mu=0$, data from~\cite{Athenodorou:2018sab}). 
Within the range of $\beta$ investigated,
the dependence of $a\sqrt{\sigma}$ on $\beta$ appears to be approximately
linear. It is, however, unlikely that this linear behavior persists
asymptotically, and it is more plausibly interpreted as a consequence of the
limited range of couplings explored in the present study. While the available
data do not allow us to draw definitive conclusions, the observed behavior of
$a\sqrt{\sigma}$ is consistent with the existence of a critical point, either
at a finite value of $\beta$ or at $\beta=\infty$.

\begin{table}[h!]
  \centering
  \begin{tabular}{l|l|l|l}
   $\beta$ & $\mu$ & $\rho_m$ & $\rho_m/(a\sqrt{\sigma})^3$ \\ \hline 
   1.8   & 0.0    & 0.015151(11)  &  0.5506(99) \\ \hline
   1.9   & 0.230  & 0.013856(22)  &  0.630(54)  \\ \hline
   2.0   & 0.4495 & 0.012240(24)  &  0.695(11)  \\ \hline
   2.1   & 0.725  & 0.012124(26)  &  0.796(13)  \\ \hline
   2.2   & 0.997  & 0.012014(29)  &  0.920(14)  \\ \hline
   2.2   & 0.0    & 0.0019668(56) &  0.4253(26) 
   \end{tabular}
   \caption{Values of the dimensionless ($\rho_m$) and dimensionful
    monopole density along the line of constant physics and, for comparison, at
   $\beta=2.2$, $\mu=0$.}
   \label{tab:monodens}
\end{table}

Finally, let us discuss whether monopoles become diluted as the
continuum limit is approached. Defining the monopole density in lattice units
as $\rho_m=N_m/L^3$ (see Eq.~\eqref{eq:Nm}), where $L$ denotes the lattice size
in lattice units, we find that $\rho_m$ decreases slightly as $\beta$ increases
along the line of constant physics (see Tab.~\ref{tab:monodens}). However, this
decrease is much slower than that observed in the $\mu=0$ theory (compare with
the results at $\beta=2.2$, $\mu=0$ in Tab.~\ref{tab:monodens}).  In
particular, the monopole density in lattice units decreases so slowly with
$\beta$ that the corresponding density in physical units,
$\rho_m/(a\sqrt{\sigma})^3$ (evaluated using the string tension values reported
in Tab.~\ref{tab:final}), actually increases along the line of constant
physics, from $\approx 0.63$ at $\beta=1.9$ to $\approx 0.92$ at $\beta=2.2$
(see Tab.~\ref{tab:monodens}).  This behavior contrasts with that of the
standard U(1) theory, where the monopole density decreases with increasing
$\beta$ even when expressed in physical units.  As previously noted, the
available numerical data are still insufficient, and our theoretical
understanding of the system is not yet developed enough to make reliable
predictions about its large-$\beta$ behavior.  In particular, we cannot exclude
the possibility that, upon further increasing $\beta$ along the line of
constant physics, the growing monopole density could eventually drive the
system into a new phase, possibly through a first-order phase transition.

\section{Conclusions}
\label{sec:concl}

The three-dimensional compact U(1) lattice gauge theory has long served as a
paradigmatic model for the study of color confinement. This model indeed
displays two complementary features: confinement can be understood analytically
within a semiclassical framework, while numerical simulations can be performed
with relatively modest computational effort. At the same time, the model
exhibits a peculiar pathology: in the continuum limit, $\beta\to\infty$, the
ratio of the glueball masses to the square root of the string tension vanishes,
see Eq.~\eqref{eq:bad}. Some consequences of this peculiar behavior have
recently been discussed in~\cite{Aharony:2024ctf,Caselle:2025vhx}.

In this work, we introduced an extension of the standard Wilson discretization
by adding to the action a term proportional to the total number of lattice
monopoles (each counted using the absolute value of its magnetic charge),
coupled to a new parameter $\mu$, see Eq.~\eqref{eq:action}. We showed that, by
tuning $\beta$ and $\mu$ simultaneously, it is possible to approach the
continuum limit along a line of constant physics, thereby removing the
pathology of the original model. More specifically, the line of constant
physics was defined by requiring the ratio
$M_{0^{--}}^{\mathrm{gs}}/\sqrt{\sigma}$ to remain constant, and we
subsequently verified that the same is true for the other low-lying mass ratios
considered in this work. Finally, we found that the string tension in lattice
units decreases as $\beta$ is increased along this trajectory, consistent with
the existence of a nonperturbative continuum limit.

Further work will surely be needed to establish the existence of a continuum
limit for this model on firmer grounds and, if such a limit exists, to
investigate its continuum properties.  Another interesting question concerns
the extent to which the semiclassical approximation remains applicable. We have
found that the slope of the line of constant physics, $\mu(\beta)$, is
remarkably close to the value predicted by the dilute monopole-gas
approximation, despite the fact that the monopole density along this trajectory
is slowly increasing. One possible explanation is that the monopole
density is still small enough for monopoles to be considered diluted, moreover
the lattice monopoles considered in this work differ in important respects from
their continuum counterparts. In particular, the magnetic charge of a lattice
monopole is bounded, $|q_{\bm x}|\le 2$ (see~\cite{DeGrand:1980eq} and
App.~\ref{sec:mono}), whereas no analogous constraint exists in the continuum.
The use of alternative lattice discretizations, like the Villain one,
could be useful to investigate whether this peculiarity of lattice monopoles is
really relevant and, more generally, to test the universality of the
construction of the line of constant physics.

\appendix

\section{Monopoles on the lattice}\label{sec:mono}

In this appendix, we briefly review, for completeness, the DeGrand-Toussaint
procedure~\cite{DeGrand:1980eq} for identifying monopoles in compact U(1)
lattice gauge theories. 

To each link variable $U_{{\bm
x},\nu}\in\mathrm{U(1)}$ we associate the phase $\phi_{{\bm
x},\nu}\in[-\pi,\pi)$, defined by
\begin{equation}
U_{{\bm x},\nu}=\exp(i\phi_{{\bm x},\nu}).
\end{equation}
The plaquette operator $P_{{\bm x},\rho\nu}$, defined in Eq.~\eqref{eq:plaq}, can then be written as
\begin{equation}
P_{{\bm x},\rho\nu}=e^{i\Phi_{{\bm x},\rho\nu}},
\end{equation}
where
\begin{equation}
\Phi_{{\bm x}, \rho\nu}=
\phi_{{\bm x},\rho}+\phi_{{\bm x}+\hat{\rho},\nu}-
\phi_{{\bm x}+\hat{\nu},\rho}-\phi_{{\bm x},\nu}\ ,
\end{equation}
and we have by construction $-4\pi \le \Phi_{{\bm x}, \rho\nu} <4\pi$.  In the
naive continuum limit, i.e. for $|\phi|\ll 1$, the quantity $\Phi_{{\bm
x},\rho\nu}$ can be interpreted as the magnetic flux through the plaquette
located at ${\bm x}$ and oriented in the $\rho\nu$ plane, as measured by a test
particle carrying unit electric charge.  It is then straightforward to show
that the total magnetic flux through the surface of an elementary lattice cube, obtained
by summing the fluxes $\Phi_{{\bm x},\rho\nu}$ associated with its six
plaquettes (with the appropriate orientations), vanishes identically.

To identify magnetic excitations associated to large values of the field we have to introduce 
the auxiliary quantity $\bar{\Phi}_{{\bm x}, \rho\nu}$, defined by 
\begin{equation}
\Phi_{{\bm x}, \rho\nu}=\bar{\Phi}_{{\bm x}, \rho\nu}+2\pi n_{{\bm x},\rho\nu}\ ,
\end{equation}
where 
$\bar{\Phi}_{{\bm x}, \rho\nu}\in [-\pi,\pi)$ and $n_{{\bm
x},\rho\nu}\in\mathbb{Z}$. 
In this way, the total magnetic flux $\Phi_{{\bm x},\rho\nu}$ is decomposed
into a physical contribution, $\bar{\Phi}_{{\bm x},\rho\nu}$, and a Dirac-string
contribution, $2\pi n_{{\bm x},\rho\nu}$. The magnetic charge $q_{\bm x}$
associated with the elementary cube whose corner is located at ${\bm x}$ and
whose edges extend along the positive lattice directions is then defined by
\begin{equation}
2\pi q_{\bm x}=\sum_{\mathrm{plaq}}\bar{\Phi}_{\rho\nu}
=-2\pi\sum_{\mathrm{plaq}}n_{\rho\nu}\ ,
\end{equation}
where the sums run over the six plaquettes bounding the cube, with the
appropriate orientations, and in the second step we used 
$\sum_{\mathrm{plaq}}\Phi_{\rho\nu}=0$.
By construction, $q_{\bm x}$ is an integer, moreover,
since it is obtained by summing six contributions, each belonging to the
interval $[-\pi,\pi)$, it follows that
\begin{equation}
-3\le q_{\bm x}<3\ .
\end{equation}
The value $q_{\bm x}=-3$ can only occur if all six plaquette angles are equal
to $-\pi$, a configuration of zero measure. Disregarding such exceptional
configurations, the allowed values of the magnetic charge are therefore $q_{\bm
x}=0,\pm 1,\pm 2$.

\section{Raw lattice data}\label{sec:rawdata}

This appendix contains, for completeness, the raw lattice data corresponding to
Tabs.~\ref{tab:res1.9}--\ref{tab:res2.2} in the main text.

\begin{table}[h!]
  \centering
  \begin{tabular}{l|l|l|l|l}
  $\mu$ & $aM_{0^{--}}^{\mathrm{gs}}$ &  
          $aM_{0^{--}}^{\mathrm{ex1}}$ &  
          $aM_{0^{++}}^{\mathrm{gs}}$  & 
          $a\sqrt{\sigma}$  \\ \hline
  0$^*$   & 0.5514(26) & 1.345(8)  & 0.978(11)  & 0.25877(28)   \\ \hline
  0.1 & 0.5956(51) & 1.42(3)   & 1.062(19)  & 0.2680(23)    \\ \hline
  0.2 & 0.6363(51) & 1.49(6)   & 1.094(22)  & 0.2774(33)    
  \end{tabular}
  \caption{Glueball masses and the string tension obtained at $\beta=1.9$ for
  several values of $\mu$, using an $18^2\times40$ lattice. For $\mu=0$, we quote
  the results of~\cite{Athenodorou:2018sab}, which were obtained on a lattice of
  the same size at this value of $\beta$.}
  \label{tab:rawres1.9}
\end{table}

\begin{table}[h!]
  \centering
  \begin{tabular}{l|l|l|l|l}
  $\mu$ & $aM_{0^{--}}^{\mathrm{gs}}$ &  
          $aM_{0^{--}}^{\mathrm{ex1}}$ &  
          $aM_{0^{++}}^{\mathrm{gs}}$  & 
          $a\sqrt{\sigma}$  \\ \hline
  0.3  & 0.5371(50) &  1.3049(99) &  0.948(17)  & 0.24743(51) \\ \hline
  0.44 & 0.5956(42) &  1.398(19)  &  1.0321(70) & 0.25892(66) \\ \hline 
  0.5  & 0.6332(29) &  1.456(11)  &  1.055(13)  & 0.26718(66) 
  \end{tabular}
  \caption{Glueball masses and the string tension obtained at $\beta=2.0$ for
  several values of $\mu$. The glueball masses were determined on a
  $18^2 \times 40$ lattice, as in~\cite{Athenodorou:2018sab}, while the string
  tension was extracted from simulations on a $12 \times 18 \times 40$ lattice.}
  \label{tab:rawres2.0}
\end{table}

\begin{table}[h!]
  \centering
  \begin{tabular}{l|l|l|l|l}
  $\mu$ & $aM_{0^{--}}^{\mathrm{gs}}$ &  
          $aM_{0^{--}}^{\mathrm{ex1}}$ &  
          $aM_{0^{++}}^{\mathrm{gs}}$  & 
          $a\sqrt{\sigma}$  \\ \hline
  0.5  &  0.4811(44) & 1.152(25) & 0.857(13)  & 0.227540(50) \\ \hline
  0.7  &  0.5631(51) & 1.337(10) & 0.980(12)  & 0.24538(64)  \\ \hline 
  0.9  &  0.6622(26) & 1.502(34) & 1.0847(93) & 0.26752(74) 
  \end{tabular}
  \caption{Glueball masses and the string tension obtained at $\beta=2.1$ for
  several values of $\mu$. The glueball masses were determined on a
  $22^2 \times 48$ lattice, as in~\cite{Athenodorou:2018sab}, while the string
  tension was extracted from simulations on a $12 \times 22 \times 48$ lattice.}
  \label{tab:rawres2.1}
\end{table}

\begin{table}[h!]
  \centering
  \begin{tabular}{l|l|l|l|l}
  $\mu$ & $aM_{0^{--}}^{\mathrm{gs}}$ &  
          $aM_{0^{--}}^{\mathrm{ex1}}$ &  
          $aM_{0^{++}}^{\mathrm{gs}}$  & 
          $a\sqrt{\sigma}$  \\ \hline
  1     & 0.5468(24)       &  1.281(14)       &  0.945(14)      & 0.23575(32) \\ \hline 
  1.1   & 0.5991(24)       &  1.347(16)       &  0.985(16)      & 0.24687(42) 
  \end{tabular}
  \caption{Glueball masses and the string tension obtained at $\beta=2.2$ for
  two values of $\mu$. The glueball masses were determined on a
  $22^2 \times 48$ lattice, as in~\cite{Athenodorou:2018sab}, while the string
  tension was extracted from simulations on a $12 \times 22 \times 48$ lattice.}
  \label{tab:rawres2.2}
\end{table}

\acknowledgments

CB and ISC acknowledge support from project PRIN 2022 ``Emerging gauge
theories: critical properties and quantum dynamics'' (20227JZKWP).  Numerical
simulations have been performed using the CSN4 cluster of the Scientific
Computing Center at INFN-PISA and the Cyclone HPC cluster at the HPC facility
of The Cyprus Institute.  It is a pleasure to thank Michele Caselle,
Alessandro Mariani and Michael Teper for useful discussions and
comments.  The idea for this work originated from discussions during the
ECT$^*$ workshop ``Bridging Analytical and Numerical Methods for Quantum Field
Theory'' (Trento, August 25-29, 2025) organised with the support of the Simons 
Collaboration on Confinement and QCD Strings. The authors thank the organizers and the
ECT$^*$ staff for their hospitality and for providing a stimulating scientific
environment, as well as the participants to the workshop for interesting
discussions.

\providecommand{\href}[2]{#2}\begingroup\raggedright\endgroup

\end{document}